\documentclass[reprint,
amsmath,amssymb,aps]{revtex4-2}
\usepackage{graphicx}
\usepackage{dcolumn}
\usepackage{bm}
\usepackage{xcolor}

\usepackage{enumitem}
\setlist{itemsep = 0.0pt}

\usepackage{booktabs}

\begin{document}

\title{LAN: A materials notation for 2D layered assemblies}

\author{Georgios A. Tritsaris$^a$}
\author{Yiqi Xie$^b$}
\author{Alexander M. Rush$^{a,+}$}
\author{Stephen Carr$^c$}
\author{Marios Mattheakis$^a$}
\author{Efthimios Kaxiras$^{a,c}$}
\affiliation{$^a$John A. Paulson School of Engineering and Applied Sciences, Harvard University, Cambridge, Massachusetts 02138, USA}
\affiliation{$^b$Institute for Applied Computational Science, Harvard University, Cambridge, Massachusetts 02138, USA}
\affiliation{$^c$Physics Department, Harvard University, Cambridge, Massachusetts 02138, USA}
\affiliation{$^+$Current address: Cornell Tech, New York, NY, USA}

\date{\today}

\begin{abstract}
Two-dimensional (2D) layered materials offer intriguing possibilities for novel physics and applications. Before any attempt at exploring the materials space in a systematic fashion, or combining insights from theory, computation and experiment, a formal description of information about an assembly of arbitrary composition is required. Here, we introduce a domain-generic notation that is used to describe the space of 2D layered materials from monolayers to twisted assemblies of arbitrary composition, existent or not-yet-fabricated. The notation corresponds to a theoretical materials concept of stepwise assembly of layered structures using a sequence of rotation, vertical stacking, and other operations on individual 2D layers. Its scope is demonstrated with a number of example structures using common single-layer materials as building blocks. This work overall aims to contribute to the systematic codification, capture and transfer of materials knowledge in the area of 2D layered materials.
\end{abstract}

\maketitle

Two-dimensional (2D) materials are an intriguing class of modern nanoscale materials which exploit physics that cannot be derived by scaling down the associated bulk structures and phenomena. As an example, we mention the transition from indirect to direct band gap in the semiconductor MoS$_2$ observed as a function of the number of stacked layers. Single-layer forms of these materials constitute nanoscopic building blocks for 2D layered assemblies \citep{geim_van_2013, ajayan_two-dimensional_2016}. Conversely, it is useful to view layered assemblies as a series of conventional crystals with weak but significant interactions between individual layers. The possibility to fabricate 2D architectures with desired combination of layers has posed fundamental questions about the properties of layered assemblies. A significant body of recent materials research has focused on the properties of single- and few-layer forms of 2D materials, including superconductivity in twisted bilayer graphene \citep{cao_unconventional_2018,lee_revealing_2014,devarakonda_evidence_2019,tritsaris_perturbation_2016,bistritzer_moire_2011,zhang_nearly_2019}. 

Combining the properties of 2D layers opens up almost unlimited possibilities for novel devices with tailor-made electronic, optical, magnetic, thermal and mechanical properties and couplings among elemental excitations. However, even with a small set of building blocks and number of layers the design space of 2D layered assemblies becomes enormous when structural features such as the relative orientation of the layers becomes important. Theory, simulation, and intelligent algorithms for automating the design and analysis of computational experiments can guide the search \citep{mounet_two-dimensional_2018,bassman_active_2018,haastrup_computational_2018}. While currently such an exploration in the lab is limited by the time required to fabricate a layered assembly, the paradigm of virtual high-throughput screening constitutes a viable approach to materials design. Specifically for the prototypical graphene, this modality of {\em in silico} materials discovery has been enabled by the recent development and use of electronic structure methods for the detailed investigation of the atomistic and electronic properties of its layered assemblies \citep{tritsaris_perturbation_2016,fang_electronic_2016,cao_unconventional_2018,carr_pressure_2018,bistritzer_moire_2011}. Combining interoperable data and information from physical and virtual experiments is expected to expedite the search.

The systematic creation, sharing and use of materials knowledge relies on devising protocols for efficient codification of materials information, in this case information about an assembly of arbitrary composition. The first step in this direction is to be able to name it, which requires a system for broadly applicable (domain-general) notation for the most trivial or the most complicated layered assemblies, existent or not-yet-fabricated. Despite the maturity of chemical language and notation such as SMILES and SMARTS for representing chemical compounds \citep{weininger_smiles_1988,d.c.i._systems_smarts_nodate}, these cannot describe critical features of a layered assembly such as the twist angle between two neighboring layers. In the computational work of Bassman {\em et al.} 2D layered assemblies were represented by strings with length equal to the number of layers based on an alphabet representing single-layer materials as distinct symbols (eg. `MoS$_2$') \citep{bassman_active_2018}. Although tailored for layered assemblies, such description does not capture defining structural elements either. 

Moreover, rather arbitrary abbreviations have been used in the literature to describe the most studied of layered assemblies: for example, abbreviations such as `TBG' and `ATMG', are often used to describe assemblies of two layers of graphene in which one layer is rotated with respect to the other (or a `twisted Bilayer Graphene') \citep{cao_unconventional_2018}, and assemblies in which the relative twists between two neighboring layers have the same magnitude but alter in sign (or `Alternating-Twist Multi-layer Graphene`) \citep{khalaf_magic_2019}, respectively. These are ambiguous with respect to the exact twist angle or the significance of it as a control parameter. In some cases, multiple abbreviations have been used interchangeably for the same material, for example, `TBG' with `tBLG'.

Here, we introduce a domain-general materials notation designed for identifying arbitrary 2D layered assemblies and composition patterns. The manuscript is organized as follows: in Section~\ref{sec:definitions} we describe the grammar underlying the notation. Section~\ref{sec:notation} demonstrates the notation using common single-layer materials as building blocks. Section~\ref{sec:implementation} briefly discusses aspects of implementation, including first-principles calculations using a local description of bonding in terms of maximally localized Wannier orbitals. 

\section{Definitions} \label{sec:definitions}

\begin{figure}
  \centering
  \includegraphics[width=\columnwidth]{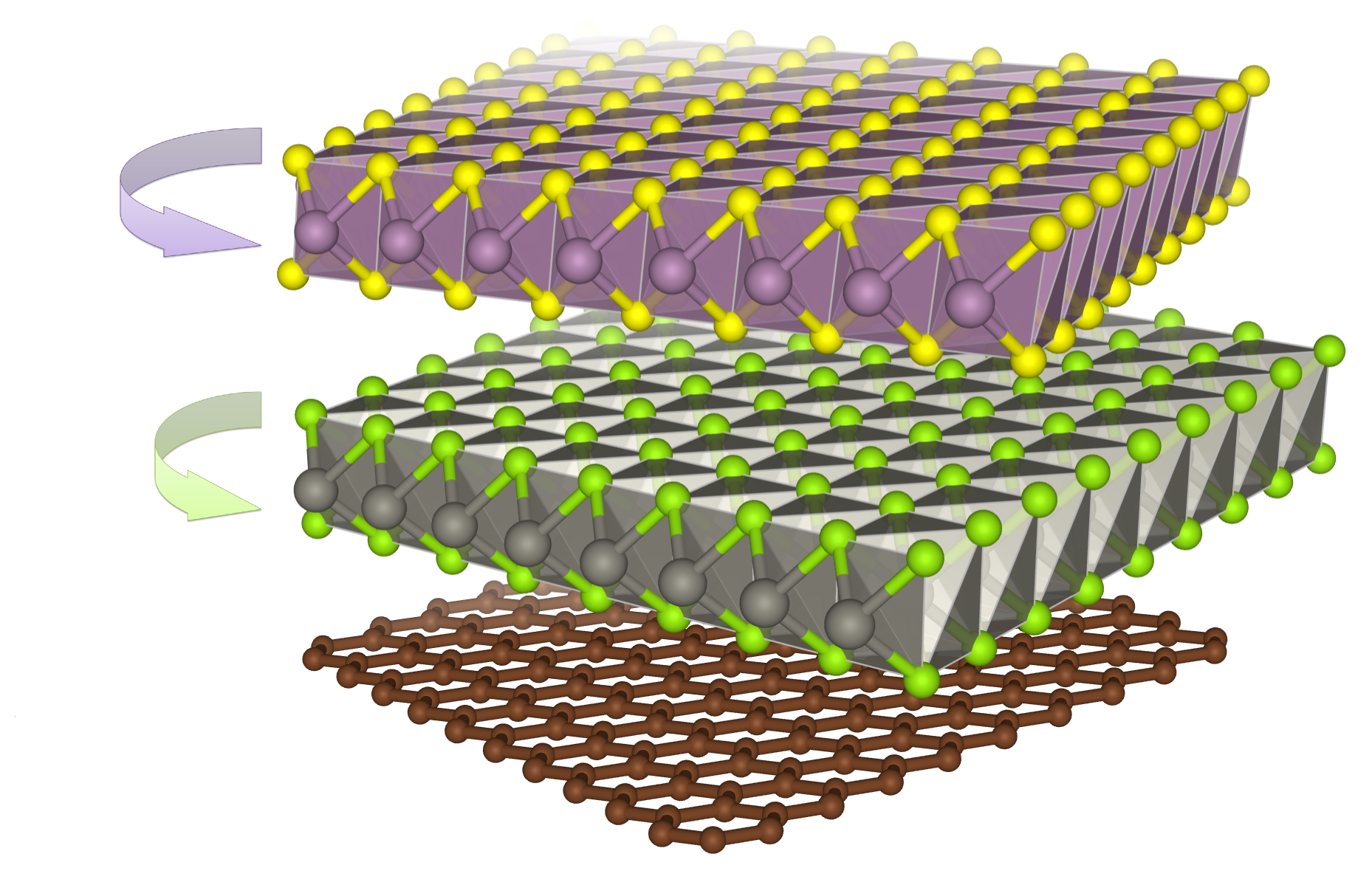}
\caption{Atomistic representation of a model twisted 2D layered assembly.}
\label{fig:assembly}
\end{figure}

The notation corresponds to a theoretical materials concept for stepwise assembly of layered structures using a sequence of rotation, vertical stacking, and other operations on individual 2D layers (Figure~\ref{fig:assembly}). Each layered assembly is treated as a design, which is described by a string that codifies the sequence of these operations. To that end, we define a context-free grammar for infix expressions to generate a language that describes the space of 2D layered materials. We use a set of production rules for replacements that closely correspond to said operations in order to define the grammar:
\begin{enumerate}
    \item S $\rightarrow$ S / S 
    \item S $\rightarrow$ S @ $\theta$
    \item S $\rightarrow$ S $>$ $a$, $b$
    \item S $\rightarrow$ S \# $a$, $b$
    \item S $\rightarrow$ (S)
    \item S $\rightarrow$ $m$
\end{enumerate}
and S as the start symbol, where $\theta \in \mathbb{R}$ denotes an angle, $(a, b)$ $\in \mathbb{R}^2$ denotes a 2D vector, and $m$ are symbols corresponding to individual layers (or building blocks) in a materials library, for example:
\begin{itemize}
    \item $m \rightarrow$ h-BN with `h-BN' the symbol for a monolayer of the insulator hexagonal boron nitride,
    \item $m \rightarrow$ G for a monolayer of the semi-metal graphene,
    \item $m \rightarrow$ G$_1$ $|$ G$_2$ $|$ G$_3$ for enumeration of an indexed materials library with three graphene flakes,
    \item $m \rightarrow$ 2H-MoS$_2$ for a monolayer of the 2H phase of the semiconductor molybdenum disulfide.
\end{itemize}
A library of atomically thin metal chalcogenides is presented in Zhou {\em et al.} \citep{zhou_library_2018} The interpretation of $m$ generally depends on context and it can describe the material, for example, as:
\begin{itemize}
    \item a crystal defined by a space group and basis,
    \item an atomistic model defined by the type and position of atoms,
    \item an exfoliated flake defined as a convex hull of a set of 2D points, or
    \item a broad class that consists of stoichiometric and defected structures, extended structural models and experimentally relevant finite flakes.
\end{itemize}

\begin{table*}
\begin{tabular}{cll} 
\toprule
Symbol & Operation & Description \\ 
\midrule
/ & Stacking & The vertical stacking of a layer or layered (sub)structure on another.\\
\midrule
@ & Rotation & The counterclockwise rotation of a layer or layered (sub)structure by \\
&& some angle about the stacking direction. \\
\midrule
$>$ & Translation & The in-plane translation of a layer or layered (sub)structure by\\
&& some 2D vector. \\
\midrule
\# & Deformation & The in-plane deformation of a layer or layered (sub)structure,\\ 
&& associated with homogeneous strain along the in-plane axes. \\
\bottomrule
\end{tabular}
\caption{Description of basic binary operations.}
\label{tab:operations}
\end{table*}

Table~\ref{tab:operations} describes basic binary operations that correspond to operations on a layer or layered (sub)structure. All operations assume the same (absolute) coordinate system. Unless otherwise stated, a Cartesian coordinate system is assumed,
as well as $\rm{\AA}$ (1~$\rm{\AA} = 10^{-10}$~m) and degrees ($360^\circ = 2\pi$) as units for expressing translation vectors and rotation angles, respectively. 

The syntax can be readily revised to explicitly assign the lowest precedence to the vertical stacking operation and the highest to the evaluation of parentheses, which we assume hereafter. Moreover, the grammar can be extended to describe layered (sub)structures more concisely using the form $n$ * $m$, where `*' corresponds to an operator of high precedence that represents a sequence of $n \in \mathbb{N}^+$ stacking operations on the same material $m$. A stitching operator `$\parallel$' can be introduced for the in-plane stitching of a layer or layered (sub)structure with another \citep{liu_-plane_2013}. Alternatively, planar combinations, or more generally any substructure, can be redefined as a new building block to simplify notation (see {\em Example 3}).

External conditions and perturbations such as pressure, temperature or a magnetic field can be used to control the properties of a layered assembly. For instance, pressurization would result in smaller interlayer distances with direct effect on their interactions and therefore properties of the assembly \citep{carr_pressure_2018}. Such effects can be accounted for as transformations on the entire assembly and they are not described by the notation.

\section{Notation} \label{sec:notation}
We demonstrate the layered assemblies notation (`LAN') with a number of example structures using three common single-layer materials as building blocks, 
\begin{center}
$m$ $\rightarrow$ G $|$ h-BN $|$ MoS$_2$.
\end{center}

{\em Example 1}: The string 
\begin{center}
MoS$_2$ / MoS$_2$ @ 3.45
\end{center}
describes the case of a bilayer of MoS$_2$ with counterclockwise relative twist angle of 3.45$^\circ$, which can be produced by applying the production rules as follows: \\
S [the start symbol]\\
$\rightarrow$ S / S 			[after application of production rule 1]\\
$\rightarrow$ S / S @ $\theta$		[rule 2]\\
$\rightarrow$ ($m$ / $m$) @ $\theta$ 	[rule 6]\\
$\rightarrow$ (MoS$_2$ / MoS$_2$) @ 3.45\\

The leftmost symbol is always the bottom layer (or substrate). Although the notation `MoS$_2$/MoS$_2$@3.45' is used here to describe a single assembly, when used in a query the same string matches not only a twisted bilayer of MoS$_2$ but also structures such as a twisted bilayer MoS$_2$ on a h-BN substrate (`h-BN/MoS$_2$/MoS$_2$@3.45'), a twisted trilayer of MoS$_2$ (`MoS$_2$/MoS$_2$@3.45/MoS$_2$@2.10'), etc. Likewise, the string `MoS$_2$/MoS$_2$@-3.45' describes a MoS$_2$ bilayer with clockwise relative twist angle. \\

{\em Example 2}: The grammar generates the string
\begin{center}
(G / G) / (G / G) @ 1.14,
\end{center}
which describes a bilayer of graphene on another bilayer after rotating the latter by 1.14$^\circ$, by applying the production rules as follows:\\
S [the start symbol]\\
$\rightarrow$ S @ $\theta$			[after application of production rule 2]\\
$\rightarrow$ S / S @ $\theta$		[rule 1]\\
$\rightarrow$ (S) / (S) @ $\theta$		[rule 5]\\
$\rightarrow$ (S / S) / (S / S) @ $\theta$ 	[rule 1]\\
$\rightarrow$ ($m$ / $m$) / ($m$ / $m$) @ $\theta$ 	[rule 6]\\
$\rightarrow$ (G / G) / (G / G) @ 1.14\\
 
{\em Example 3}: The notation `G/G' can be used to substantiate the common abbreviation `BLG' for a bilayer of graphene with the production rules 
\begin{center}
$m \rightarrow$ BLG and BLG $\rightarrow$ G / G.
\end{center}
When treating the stacking order of layers explicitly is important, the translation operation is used to describe the relative shift. For instance, the common AB-stacking is described by such production rules as
\begin{center}
$m \rightarrow$ BLG$\rm{_{AB}}$, and 
BLG$\rm{_{AB}}$ $\rightarrow$ G / G $>$ $a\rm{_{AB}}$, $b\rm{_{AB}}$ or\\
BLG$\rm{_{AB}}$ $\rightarrow$ G$\rm{_{A}}$ / G$\rm{_{B}}$; 
G$\rm{_{B}}$ $\rightarrow$ G$\rm{_{A}}$ $>$ $a\rm{_{AB}}$, $b\rm{_{AB}}$,
\end{center}
with $(a\rm{_{AB}}, b\rm{_{AB}})$ the translation vector from AA- to AB-stacking order. 
Likewise, the production rules 
\begin{center}
$m \rightarrow$ tBLG and tBLG $\rightarrow$ G / G @ $\theta$
\end{center}
associate the abbreviation `tBLG' with a twisted bilayer of graphene. 

\section{Implementation} \label{sec:implementation}
\begin{figure}
  \centering
  \includegraphics[width=\columnwidth]{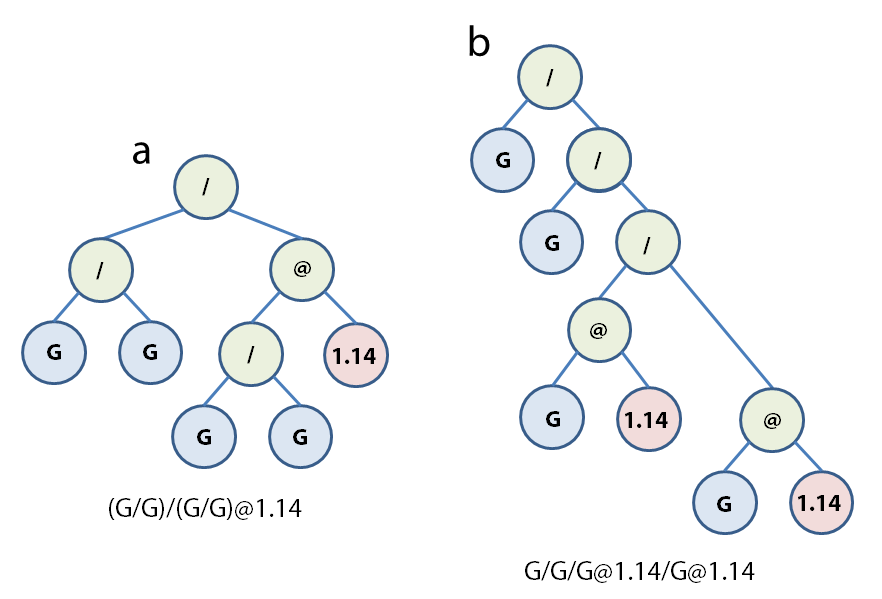}
\caption{Binary expression trees for a twisted 4-layer assembly of graphene (`G'). The green, blue, and red fill colors describe the operators (vertical stacking or rotation), single-layer material (graphene), and twist angle, respectively.}
\label{fig:tree}
\end{figure}

In terms of a computational procedure, the problem of parsing a string is to find a binary expression tree for that string describing the operations in constructing a layered assembly. In many cases a number of equally valid strings can be written for one assembly. For instance, the string `(G/G)/(G/G)@1.14' can be rewritten as `G/G/G@1.14/G@1.14'. The two binary expression trees corresponding to the twisted 4-layer assembly of graphene are shown in Figure~\ref{fig:tree}. Devising an algorithm for a canonical representation is beyond the scope of this work, nevertheless one approach would rely on such an expansion (distribution of operations) to a list of layers with associated operations as attributes. 
 
Conversely, a string is obtained by printing the symbol associated with the nodes encountered in a traversal of the expression tree. By reducing one node at the time, we gradually progress towards the tree root until when only the root node remains (the whole process can also be run recursively starting from the root node). Transversal of the graph can be used, for example, to:
\begin{itemize}
    \item convert a given string into a 2D or 3D model of the corresponding layered assembly, 
    \item prepare input for a physical or virtual measurement, or
    \item instruct a robotic arm in constructing the physical structure. 
\end{itemize}    

{\em Application 1}: Imagine a synthesis tool that enables rapid and precise fabrication of layered materials by assembling the building blocks in a controlled manner like a press. The sequence of assembly instructions associated with Figure~\ref{fig:tree}b are:
\begin{enumerate}
    \item Retrieve a monolayer of graphene
    \item Stack layer
    \item Retrieve a monolayer of graphene 
    \item Stack layer
    \item Retrieve a monolayer of graphene
    \item Rotate layer by 1.14$^\circ$ counterclockwise
    \item Stack layer
    \item Retrieve a monolayer of graphene
    \item Rotate layer by 1.14$^\circ$ counterclockwise
    \item Stack layer
    \item Stop
\end{enumerate}

A tree-based representation of layered assemblies also makes evolutionary approaches and related global search algorithms suitable for exploring the materials space \citep{johannesson_combined_2002}. Although a list-based representation in some cases might be preferable for implementation \citep{bassman_active_2018}, the underlying grammar still provides the means for consistent interpretation and systematic refinements and derivations.\\

{\em Application 2}: Adopting a functional expression for scientific workflows,
it is written for the band gap, E$\rm{_g}$, of G/G for instance, 
\begin{center}
E$\rm{_g}$[G/G] = 0.
\end{center}
Within the context of materials modeling the above functional expression essentially describes the output of a sequential computational workflow linking the preparation of a structural model for bilayer graphene, the calculation of its electronic structure, and post-processing to obtain its band gap. 

\begin{figure*}
  \centering
  \includegraphics[width=\textwidth]{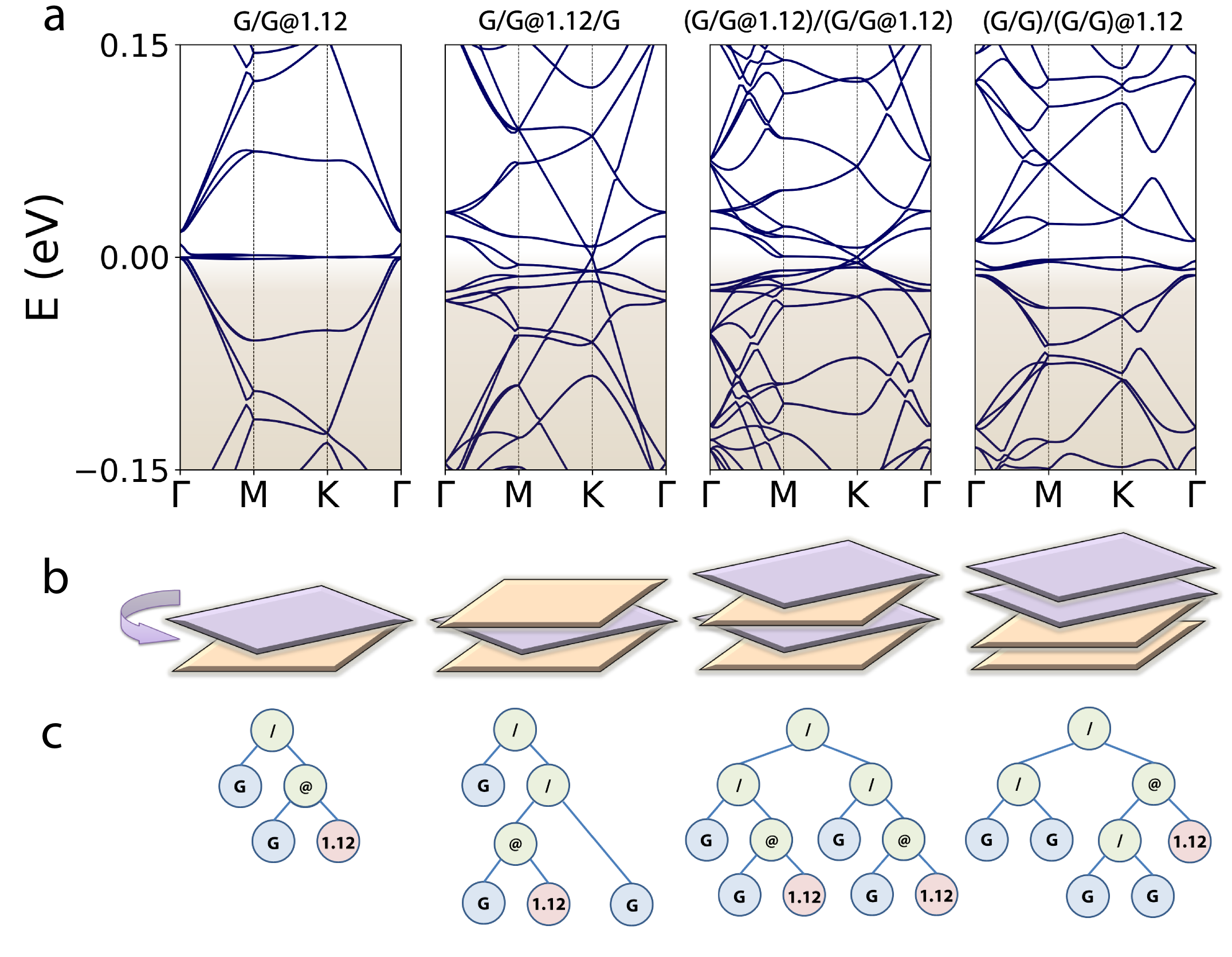}
\caption{a) Low-energy band structures for four twisted ($\theta = 1.12^\circ$) layered assemblies of graphene (`G'), obtained using a tight-binding model. The shaded area marks occupied electronic states. b) Schematic of layered assemblies with exaggerated twist angle. c) Corresponding binary expression trees (color code follows that of Figure~\ref{fig:tree}).}
\label{fig:bs}
\end{figure*}

When one layer is rotated by about 1.1$^\circ$, commonly referred to as the `magic angle', bilayer graphene becomes an insulator or a superconductor with flat electronic bands in the single-particle band structure. These low-dispersion bands serve as indicators of this emergent physical behavior \citep{cao_unconventional_2018}. For a practical demonstration, we use next a computational workflow based on a first-principles tight-binding model to calculate the low-energy band structures of selected two-, three-, and four-layer assemblies of extended graphene layers with the same twist ($\theta = 1.12^\circ$), shown in Figure~\ref{fig:bs}. These layered assemblies are:
\begin{enumerate}
    \item G/G@1.12: a twisted bilayer of graphene (often referred to in the literature as `tBLG' or `TBG') \citep{cao_unconventional_2018,bistritzer_moire_2011,tritsaris_perturbation_2016},
    \item G/G@1.12/G: a twisted sandwitched graphene (`TSWG') \citep{carr_coexistence_2019,khalaf_magic_2019}, 
    \item (G/G@1.12)/(G/G@1.12): an alternating-twist multi-layer graphene (`ATMG') \citep{khalaf_magic_2019}, and
    \item (G/G)/(G/G)@1.12: a twisted double bilayer of graphene (`TDBG', `TBBG') \citep{chebrolu_flat_2019,shen_observation_2019},  
\end{enumerate} 
which we distinguish and discuss, unambiguously, using LAN. 

The construction of the tight-binding Hamiltonian relies on using maximally localized Wannier functions for obtaining an exact representation of density functional theory wavefunctions in a basis set of localized orbitals  \citep{marzari_maximally_2012,fang_electronic_2016}. A fully parametrized model for in-plane and inter-plane p$_z$ orbital interactions in obtained, which is then solved by efficient diagonalization of the system’s Hamiltonian. A detailed presentation of the used method is provided in Fang {\em et al.} \citep{fang_ab_2015,fang_electronic_2016}. We use 60 points to sample the Brillouin zone along the direction connecting the high-symmetry k-points $\Gamma$, M, and K. The first step of the workflow entails the parsing of a given string-based description into a simple list-based representation for rotations (either 0 or $\theta$) to inform the construction of the Hamiltonian to be solved. A tree-based representation of each calculated structure is also provided in Figure~\ref{fig:bs}c. 

Although our calculations do not account for atomic relaxation, they demonstrate the non-trivial effect of the atomic environment (neighboring layers) of a twisted bilayer of graphene, G/G@1.12, on its electronic structure. In all cases the flat electronic bands are distorted as more layers are added while keeping $\theta$ fixed: compare for instance the electronic states near the Fermi level (0 eV) between G/G@1.12 and G/G@1.12/G in Figure~\ref{fig:bs}a. Both G/G@1.12/G and (G/G@1.12)/(G/G@1.12) host electronic bands which are more disperse because the alternating layers of graphene effectively enhance interlayer coupling and the `magic angle' instead occurs around $1.5^\circ$ \citep{carr_coexistence_2019,khalaf_magic_2019}. A distinguishing feature of G/G@1.12/G is a Dirac cone at the K-point that is very similar to this of an isolated graphene layer. A detailed investigation of the electronic structure of graphene-based layered assemblies is nevertheless beyond the scope of this work. 

\section{Conclusions} \label{sec:conclusions}
We introduced a notation based on a grammar that can be used to describe the space of 2D layered materials from monolayers to twisted assemblies of arbitrary composition, existent or not-yet-fabricated, and support the integration of information from first-principles calculations to fabrication and characterization. Each layered assembly is treated as a design, which is described by a string that codifies a sequence of vertical stacking and other operations on 2D building blocks. Using the notation, common abbreviations in the pertinent literature are substantiated. 

Most importantly, this work contributes to the systematic codification, capture and transfer of materials knowledge in the area of 2D layered materials. The notation relies on rigorous definitions which can be refined in a systematic fashion. Being domain-generic, it facilitates the integration of information from theory, computation and experiment, while derivations tailored for particular knowledge domains, for instance, theoretical modeling or synthetic work, should improve its descriptive power within these domains.

Finally, calculations of selected twisted layered assemblies of graphene based on a tight-binding model demonstrate the non-trivial effect that the atomic environment can have on a neighboring layer. This invites a high-throughput approach to systematically search the materials space, and the use of flexible notation such as LAN for cataloguing it. Such an extended and detailed investigation is the focus of on-going work.

\section*{Acknowledgements}
We thank the following for their useful feedback:
Philip Kim,
Daniel T. Larson, 
Andrew Pierce,
Di S. Wei,
Amir Yacoby,
Ziyan Zhu at Harvard University; 
Joseph G. Checkelsky,
John Ingraham,
Daniel Rodan-Legrain,
Takehito Suzuki, 
Kenjie Yasuda at MIT; and 
Kevin Yager at BNL.

This work was supported in part by the U.S. Department of Energy, Office of Science, Basic Energy Sciences, under Award No. DE-SC0019300. We used computational resources of the National Energy Research Scientific Computing Center (NERSC), a DOE facility operated under Contract No. DE-AC02-05CH11231.

\bibliography{References}

\end{document}